\documentclass[12pt]{article}
\usepackage{epsfig}
\usepackage{graphics}
\usepackage{psfrag}
\usepackage{amsmath,amssymb}

\newcommand{\beq}{\begin{equation}}
\newcommand{\eeq}{\end{equation}}
\newcommand{\bea}{\begin{eqnarray}}
\newcommand{\eea}{\end{eqnarray}}

\newcommand{\e}{\mbox{e}}

\newcommand{\g}{\gamma}

\newcommand{\lam}{\lambda}
\newcommand{\La}{\Lambda}
\newcommand{\Lam}{\Lambda}
\renewcommand{\b}{\beta}
\renewcommand{\a}{\alpha}
\newcommand{\n}{\nu}
\newcommand{\m}{\mu}

\newcommand{\del}{\delta}
\newcommand{\Del}{\Delta}
\newcommand{\sg}{\sigma}

\newcommand{\oh}{\frac{1}{2}}

\newcommand{\ra}{\rangle}
\newcommand{\la}{\langle}
\newcommand{\prt}{\partial}
\newcommand{\mi}{\!-\!}

\newcommand{\cD}{{\cal D}}

\newcommand{\cO}{{\cal O}}

\newcommand{\hg}{{\hat{g}}}

\newcommand{\rf}[1]{(\ref{#1})}
\newcommand{\figref}[1]{Fig.\ \ref{#1}}

\newcommand{\tri}{\triangle}

\begin{document}

\begin{center}
\vspace{24pt}

{ \large \bf 
Two-Dimensional Quantum Geometry\footnote{Lectures 
presented  at ``The 53rd Cracow School of Theoretical Physics: 
Conformal Symmetry and Perspectives in Quantum and Mathematical Gravity'', 
June 28 - July 7, 2013, Zakopane, Poland}}

\vspace{30pt}

{\sl J. Ambj\o rn}$\,^{a,b}$
and 
{\sl T.Budd}$\,^{a}$

\vspace{48pt}
{\footnotesize

$^a$~The Niels Bohr Institute, Copenhagen University\\
Blegdamsvej 17, DK-2100 Copenhagen \O , Denmark.\\
{email: ambjorn@nbi.dk, acipsen@gmail.com}\\

\vspace{10pt}

$^b$~Institute for Mathematics, Astrophysics and Particle Physics (IMAPP)\\ 
Radbaud University Nijmegen, Heyendaalseweg 135,
6525 AJ, Nijmegen, The Netherlands. 

}

\vspace{96pt}
\end{center}

\begin{center}
{\bf Abstract}
\end{center}

\noindent
In these lectures we review our present understanding of
the fractal structure of two-dimensional Euclidean quantum gravity
coupled to matter.

\vspace{12pt}
\noindent

\vspace{24pt}

\newpage

\section{Introduction}\label{intro}

A noble task in ancient, pre-AdS/CFT time was 
to find a non-perturbative definition of Polyakov's bosonic
string theory. The formal partition function was defined 
by the path integral: 
\beq\label{1.1}
Z = \int \cD [g_{\a\b}] \;\e^{-\Lam \int d^2\!\xi \sqrt{g}} \int \cD_g X_\m \;
\e^{-\frac{1}{2\alpha'}\int d^2\!\xi 
\sqrt{g} g^{\a\b} \prt_\a X_\m \prt_\b X_\m}.
\eeq
Here
$[g_{\a\b}]$ represents a {\it continuous} 2d geometry of some fixed topology. Assume that the 
set of piece-wise linear geometries one can obtain by gluing together 
equilateral triangles with  link length $a$ is uniformly dense
in the set of continuous 2d geometries when $a \to 0$. 
Each such geometry can be identified with an abstract triangulation.
By placing the matter field $X_\m(\xi)$ in the center of each triangle
and using the natural discretized version of the matter Lagrangian 
in \rf{1.1} we obtain a lattice regularization of the 
action, for which the lattice spacing $a$ acts as a UV cut-off. Summing over
the abstract triangulations provides a lattice regularization of the integral 
over geometries in \rf{1.1}, coined {\it Dynamical Triangulations} (DT) \cite{david,kkm,adf}.
If the assumption about the denseness of these triangulations
in the set of continuous geometries holds, we expect to obtain the 
continuum path integral in the limit $a \to 0$. Of course, it is 
to be expected that one has to renormalize the bare coupling 
constants entering in the lattice partition function to 
recover the continuum results. If we work in units where the 
lattice spacing $a$ is put to one, we obtain the
dimensionless DT partition function
\beq\label{1.2}
Z(\mu) = \sum_T \e^{-\m N_T } 
\int'  \Big(\prod_{\tri\in T}\prod_{\n=1}^d dx_{\n}(\tri)\Big) \;
\e^{-\oh\sum_{\tri,\tri'} (x_\n(\tri) -x_\n(\tri'))^2}
\eeq
for the bosonic string, where the overall sum is over triangulations $T$ with $N_T$ triangles and the sum in the exponent is over pairs $\tri, \tri'$ of neighboring triangles.

\subsection{The free particle}

To understand how to obtain the continuum limit 
of \rf{1.2}, it is 
useful to study the simpler system of a free particle. 
In this case the propagator $G(X_\n,X'_\n)$ has the
path integral representation
\beq\label{1.3}
G(X_\n,X'_\n) = \int \cD [g] \e^{-\Lam \int d\xi \sqrt{g} }
\int \cD_g X_\n \; \e^{- \frac{1}{2\a'}\int_0^1 d\xi \sqrt{g} g^{-1} 
(\prt_\a X_\n)^2},
\eeq
where $X_\n(0)=X_\n$ and $X_\n(1)=X'_\n$, and
$[g]$ is the geometry of a world line, i.e. $d\ell^2 = g(\xi)d \xi^2$
and $\int d\xi \sqrt{g} = \ell $. The structure of eq.\ \rf{1.3} is 
quite similar to that of eq.\ \rf{1.1}. The path integral is discretized
by dividing the worldline in $n$ equal steps (the equivalent of 
the equilateral triangles for DT) and using dimensionless variables:
\beq\label{1.4}
G(x_\n,x'_\n,\m) = \sum_n \e^{-\mu n} \int \Big(
\prod_{i=1}^n\prod_{\n=1}^{d} dx_\n(i)\Big) \; 
\e^{-\oh\sum_{i=1}^n (x_\n(i) -x_\n(i-1))^2},
\eeq
with $x(0)=x$ and $x(n)=x'$. One can perform the Gaussian integrations:
\beq\label{1.5}
\int\Big(
\prod_{i=1}^n\prod_{\n=1}^{d} dx_\n(i)\Big) 
\e^{-\oh\sum_{i=1}^n (x_\n(i) -x_\n(i-1))^2} = 
\frac{(2\pi)^{nd/2}}{(2\pi n)^{d/2}} \; \e^{-\frac{ (x_\n-x'_\n)^2}{2 n}}.
\eeq
Introducing $\m_c= \oh d\log (2\pi)$, we get
\beq\label{1.6}
G(x_\n,x'_\n,\m) = \sum_n \frac{1}{(2\pi n)^{d/2}}
\;\e^{-(\mu-\mu_c) n}\; \e^{-\frac{ (x_\n-x'_\n)^2}{2 n}},
\eeq
leading to
\beq\label{1.7}
G(x_\n,x'_\n,\m) \approx f(|x_\n-x'_\n |)\; \;\e^{-m(\m)|x_\n-x'_\n|},~~~~~~
m(\m) \propto \sqrt{\m-\m_c}.
\eeq
Performing a {\it mass renormalization} and a scaling,
\beq\label{1.8}
m^2(\m)= \m-\m_c =m^2_{\textrm{ph}} a^2, ~~~x \, a = X,~~x'\,a = X',
~~t=na^2
\eeq
we obtain the standard proper time representation of the free 
relativistic propagator
\beq\label{1.9}
G(X_\n,X'_\n;m_{\textrm{ph}})=\!\! \lim_{a\to 0} a^{2-d} G(x_\n,x'_\n,\m) =\!\!
\int_0^\infty\!\!\!\! \frac{dt}{(2\pi t)^{d/2}}\; 
\e^{-m_{\textrm{ph}}^2 t-\frac{(X_\n-X'_\n)^2}{2t}}. 
\eeq

\begin{figure}
\centerline{{\scalebox{0.5}{\rotatebox{0}{\includegraphics{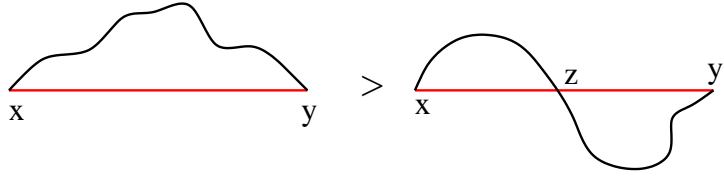}}}}}
\caption{Decomposition of random walk into two random walks}
\label{fig1}
\end{figure}

The explicit, well-defined path integral representation \rf{1.4} 
of the free particle is useful for analyzing simple basic 
properties of the propagator. Let us just mention one such 
property, the exponential decay of the propagator for large distances.
Why can the propagator not fall of faster than exponentially at large 
distances? The answer is found by looking at \figref{fig1}. The set 
of paths from $x$ to $y$ has as  a subset the set of paths intersecting
the straight line connecting $x$ and $y$ at a point $z$. 
A path in this subset is a union of a path from $x$ to $z$ and from 
$z$ to $y$. Since the action for such a path is the sum 
of the actions of the path from $x$ to $z$ and the path 
from $z$ to $y$, it is not difficult to show
\beq\label{1.10}
G(x,y) \geq G(x,z) G(z,y),
\eeq
 i.e.
\beq\label{1.10a}
-\log G(x,y) \leq -\log G(x,z) -\log G(z,y).
\eeq
The {\it subadditivity} of $-\log G(x,y)$ implies that there exits
a positive constant $m$ such that 
\beq\label{1.11} 
-\log G(x,y) \sim m |x-y|~~~{\rm for}~~~|x-y| \to \infty,
\eeq  
i.e.\
\beq\label{1.12}
G(x,y) \sim \e^{-m|x-y|}~~~{\rm for}~~~|x-y| \to \infty.
\eeq
The constant $m$ is the mass of the particle
(which can be zero in special cases). 
 
\subsection{The bosonic string}

One can also perform the Gaussian integration in the string case:
\beq\label{1.13}
\int'\!\! \Big(\!\prod_{\tri\in T_N}\prod_{\n=1}^d dx_{\n}(\tri)\Big)
\e^{-\oh\sum_{\tri,\tri'} (x_\n(\tri) -x_\n(\tri'))^2}\! = \!
\Big(\det (-\Del^{'}_{T_N})\Big)^{-d/2}\!,
\eeq
where $\Del_{T_N}$ is the combinatorial Laplacian on the dual $\phi^3$-graph.
The prime indicates that the constant zero mode is 
projected out in the determinant. We find
\beq\label{1.14}
Z(N) = \sum_{T_N} \Big(\det (-\Del^{'}_{T_N})\Big)^{-d/2} = 
\e^{\m_c N} N^{\g(d)-3}\Big(1+ \cO\Big(\frac{1}{N^2}\Big)\Big) 
\eeq
and
\beq\label{1.15}
Z(\mu) = \sum_N \e^{-\m N} Z(N) = \sum_N \e^{-(\m-\m_c)N} N^{\g(d)-3} 
\Big(1+ \cO\Big(\frac{1}{N^2}\Big)\Big).
\eeq
In the scaling limit $\m \to \m_c$ one may identify
\beq\label{1.16}
\m -\m_c = \Lam a^2,~~~
(\m-\m_c)N_T = \Lam \int d^2\xi \sqrt{g}.
\eeq

Equation \rf{1.15} is valid for geometries with fixed topology of the sphere, 
but $Z(\mu)$ generalizes naturally to surfaces with $n$ 
boundaries $\{\gamma_i\}$ of fixed length $L_i$ on which the 
coordinates $x_\mu$ are fixed.
In particular, in the limit $L_i\to 0$ we obtain the $n$-point function 
$G(x_1,\ldots,x_n;\m)$ for spherical string world sheets 
with $n$ marked points at prescribed positions $x_1,\ldots,x_n$.

\begin{figure}
\centerline{{\scalebox{0.3}{\rotatebox{0}{\includegraphics{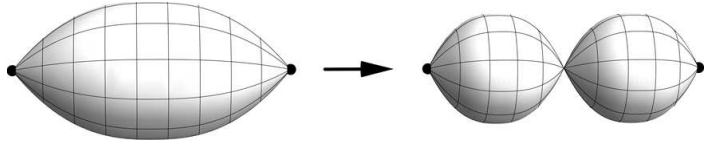}}}}}
\vspace{12pt}
\caption{Subadditivity of the string two-point function.}
\vspace{12pt}
\label{fig2}
\end{figure}

A basic property of the two-point function $G(x_1,x_2;\mu)$ is subadditivity.
The argument is essentially the same as for the particle, 
except that random surfaces are involved 
instead of random walks, as illustrated in \figref{fig2}.
Therefore we find
\beq\label{1.17}
  G(x_1,x_2;\m) \sim \e^{-m(\m)|x_1-x_2|},~~~~~m(\m) \geq 0.\
\eeq

\begin{figure}
\centerline{{\scalebox{0.28}{\rotatebox{0}{\includegraphics{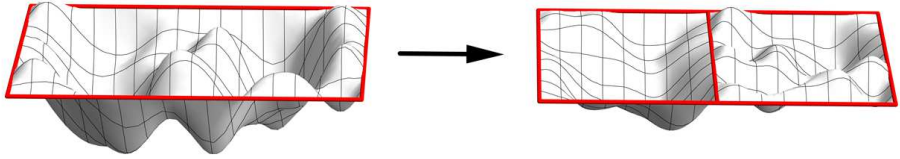}}}}}
\caption{Subadditivity of the Wilson loop.}
\label{fig3}
\end{figure}

Similarly we may consider the planar ``Wilson loop'' $G(\g_{L_1\times L_2},\m)$, 
corresponding 
to the partition function with one boundary $\g_{L_1\times L_2}$ of 
length $2L_1+2L_2$ corresponding to a 
rectangular loop in $\mathbb{R}^d$ with sides of length $L_1$ and $L_2$. 
As illustrated in \figref{fig3}, 
$G(\g_{L_1\times L_2},\m)$ is subadditive both in $L_1$ and $L_2$, 
and therefore we 
obtain\footnote{For a more precise argument see \cite{book}, section 3.4.4.}
\beq\label{1.18}
G(\g_{L_1\times L_2},\m) \sim \e^{-\sg(\m) A(\g_{L_1\times L_2})},
~~~~\sg(\m) \geq 0,
\eeq
where $A(\g_{L_1\times L_2}) = L_1 L_2$ is the area of the loop, and $\sigma(\mu)$ is known as the string tension.

\begin{figure}
\centerline{\scalebox{0.35}{\rotatebox{0}{\includegraphics{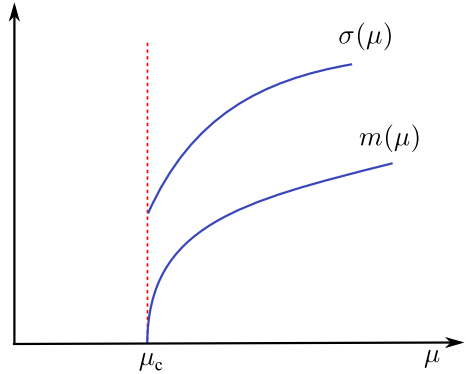}}}}
\caption{The scaling of the bare mass and the bare string tension as a 
function of the bare coupling constant $\m$.}
\label{fig4}
\end{figure}

However, the dominant worldsheet surfaces look completely different
from the nice surfaces shown in \figref{fig2} and \figref{fig3}.
The reason for this is shown in \figref{fig4}. 
It is seen from the figure that, while the mass of the two point function
scales to zero at a critical point, which is needed if one 
wants a continuum limit, this is not the case for the string 
tension $\sigma(\mu)$ (\cite{ad} or \cite{book}, theorem 3.6).
The consequence is that the physical string tension 
scales to infinity as $\mu\to\mu_c$:
\beq\label{1.19} 
m(\m) = (\m-\m_c)^\n = m_{\textrm{ph}}\, 
a^\n,~~~~\sg(\m)=\sg_{\textrm{ph}}\, a^{2\n},~~~
\sg_{\textrm{ph}} \to \infty.
\eeq

\begin{figure}[t]
\centerline{\scalebox{0.35}{\rotatebox{0}{\includegraphics{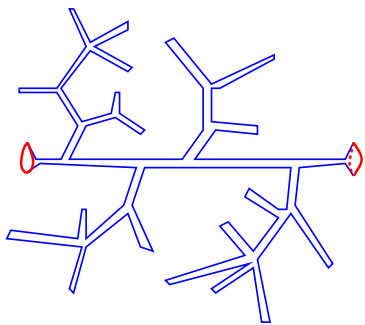}}}}
\caption{Branched polymer surfaces dominate the 
bosonic string two-point function.}
\label{fig5}
\end{figure}

An infinite string tension implies that any surface with finite area is 
forbidden unless it is dictated by some imposed boundary conditions.
A typical surface with no area contributing to the 
two-point function $G(x_1,x_2,\m)$ is shown in \figref{fig5}. 
Such surfaces are called branched 
polymer (BP) surfaces.
They have only {\it  one} mass excitation corresponding to a free particle,
since one basically obtains a random walk representation corresponding 
to the free particle by scaling away the branches decorating the 
shortest path from $x_1$ to $x_2$ for a given surface 
connecting $x_1$ and $x_2$.

\begin{figure}[t]
\centerline{\scalebox{0.4}{\rotatebox{0}{\includegraphics{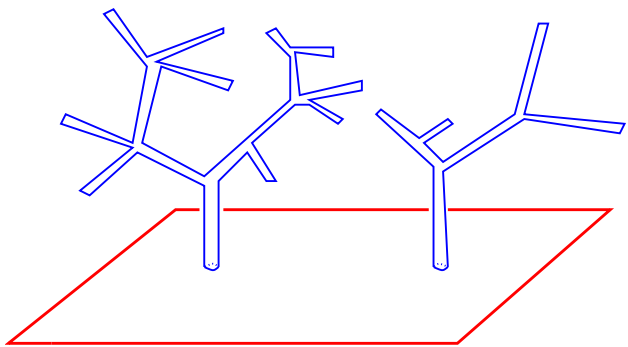}}}}
\caption{The fluctuations around the minimal surface in the path integral of the Wilson loop are of the form of branched polymers.}
\label{fig6}
\end{figure}

In the case of the Wilson loop we are summing over surfaces where 
the boundary is fixed. Therefore we have a minimal-area surface stretching 
to the boundary. The fluctuations around this surface, however, 
are again branched polymers, as shown in \figref{fig6}, 
and are nothing like the surface in \figref{fig3}. 

The conclusion is that the bosonic string theory defined through
a regulated path integral where all surfaces have positive 
weight does not exist. The reason that we do not obtain the 
standard bosonic string, 
despite such a well-defined procedure, is that 
the two-point function of the standard bosonic string has 
tachyonic mass excitations, which are excluded by our construction
and which make standard bosonic string theory sick.

\section{Non-critical string theory}

However, interpreting the string world sheet as 2d space-time, we can view
Polyakov's bosonic string theory in $d$ dimensions as 2d gravity coupled to 
$d$ massless scalar fields, i.e. to a conformal field theory with central 
charge $c=d$. Therefore, as another route towards the bosonic string,
we can study 2d quantum gravity coupled to (conformal) field theories.
Surprisingly this theory, called non-critical string 
theory, has a rich structure as long 
as the central charge $c \leq 1$. 

The regularized version of such a theory is typically obtained as 
follows: assume we have a conformal field theory originating
from a field theory on a regular lattice. Usually the lattice 
theory has a critical point with a second-order phase 
transition and the continuum conformal field theory is then 
defined at the critical point. This lattice field theory can 
usually be transfered from a regular lattice to
a random one, hence, also to the random lattice appearing in
the DT formalism. 

Including a summation over different lattices
in ensemble averages
is what is called an annealed average in the context of 
condensed matter physics. Here it will play the role of 
integrating over 2d geometries, as for the bosonic string. 
  
The partition function of 2d gravity coupled to matter can be written as
\beq\label{2.1}
Z = \sum_N \e^{-\m N} \sum_{T_N} Z_{T_N} ({\rm matter}),
\eeq
where $Z_{T_N} ({\rm matter})$ is the matter partition function 
on a fixed triangulation $T_N$.
A typical example is the Ising model coupled to DT \cite{kazakov}, 
\beq\label{2.3}
  Z_{T_N}(\b) = 
\sum_{\sg_\tri=\pm 1} \exp\Big[\b \sum_{\tri,\tri'}\; \sg_\tri \sg_{\tri'}\Big].
\eeq
The partition function scales as
\beq\label{2.4}
Z_N(\b)=\sum_{T_N} Z_{T_N} (\b) = 
\e^{\m_c(\b) N}\, N^{\g(\b)-3}\Big(1+O(N^{-2})\Big),
\eeq
\beq\label{2.5} 
Z(\b) = \sum_N \e^{-\m N} Z_N(\b)= 
\sum_N \e^{-(\m-\m_c(\b))N}N^{\g(\b)-3}(1+\cdots).
\eeq
Here $\m_c(\b)$ appears as the critical ``cosmological'' constant
for the geometries, such that one obtains universes with 
infinitely many triangles when $\m \to \m_c$ from above. 
This is similar to the situation for the free particle and 
the bosonic string and we clearly want to take that limit 
in order to recover continuum physics from the lattice theory. 
However, it also follows from \rf{2.4} that 
it has the interpretation as the the free energy density  of spins
in the annealed ensemble. 

The model has 
a phase transition at a critical $\b_c$, the transition being 
third order rather than the standard second order 
phase transition \cite{kazakov}.
At the transition point $\g(\b)$
jumps from -1/2 to -1/3. 
The interpretation is as follows: on a regular lattice 
the Ising spin system also has a phase transition at 
a certain critical temperature $\b_c$. The transition is 
a second order transition and at the transition point 
the spin system describes the 
continuum conformal field theory of central charge $c=1/2$. 
The lattice theory, defined on the annealed average of 
lattices, describes at its critical point the $c=1/2$ 
conformal field theory coupled 
to 2d quantum gravity, the average over the DT lattices being the 
path integral over geometries. It is not surprising that 
the transition can change from a second order to a third order 
transition, the randomness of the lattices and the averaging over 
different lattices making it more difficult to build up large 
critical spin clusters at the phase transition point.
Maybe it is more surprising that there is a transition at all.
But it is known to be the case, since one can solve the model 
analytically. One finds that the critical spin exponents have 
changed compared to Onsager exponents on a regular lattice.
Thus the continuum conformal field theory has changed due to 
the interaction with 2d quantum gravity. Further, as we 
mentioned, the exponent $\g(\b)$ jumps at $\b_c$. The 
exponent $\g(\b)$ as it appears in \rf{2.5} reflects average fractal geometric
properties of the ensemble of random geometries appearing in the path 
integral. Thus a change in the exponent reflects that the 
conformal field theory back-reacts on the geometry 
and changes its fractal properties, something we will discuss 
in detail below.  Away from $\b_c$ the Ising model 
is not critical, and the lattice spins couple only 
weakly to the lattice. For all $\b \neq \b_c$ one has $\g(\b) = -1/2$
 and this can then be viewed as
the exponent for ``pure 2d Euclidean gravity'' without matter fields.

\subsection{Continuum formulation}
 
One can study 2d quantum gravity coupled to matter fields 
entirely in the continuum. Just like for the partition function 
\rf{1.1} for the bosonic string, we can write formally
\beq\label{2.6}
  Z = \int \cD [g_{\a\b}]\;\e^{-\Lam A(g)} \int \cD_g \psi
\;\e^{-S(\psi,g)},~~~~A(g) = \int d^2 \xi \,\sqrt{g}, 
\eeq
where $\psi$ represents some matter field.
A partial gauge fixing, to the so-called conformal gauge 
$g_{\a\b} = \e^{\phi}\hg(\tau_i)$ leads to  
\beq\label{2.7}
Z(\hg) = 
 \int \cD_{\hg} \phi\, \e^{-S_L(\phi,\hg)},
\eeq
where $S_L(\phi,\hg)$ is fixed by the 
requirement that $Z(\hg)$ is independent of $\hg$, namely \cite{DDK}
\beq\label{2.8}
S_L(\phi,\hg) = \frac{1}{4\pi} \int d^2 \xi \sqrt{\hg} \; 
\Big( (\prt_\a \phi)^2 + Q\, \hat{R}\,\phi +\mu \,\e^{2\b \phi} \Big),
\eeq
\beq\label{2.9}
Q = \sqrt{(25-c)/6},~~~~Q = 1/\b + \b.
\eeq

\begin{figure}
\centerline{\scalebox{0.5}{\rotatebox{0}{\includegraphics{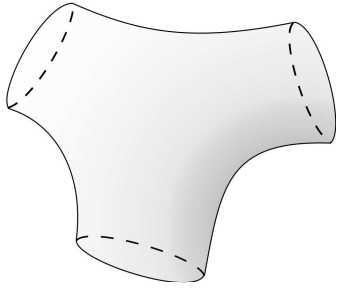}}}}
\caption{The 3-loop function}
\label{fig7}
\end{figure}

Even for $c=0$ we have a non-trivial theory.
The $c=0$ partition function  can be obtained  explicitly  at the  regularized 
level simply  by counting the triangulations, since there are no 
matter fields. A slightly non-trivial structure 
can be imposed by $n$ boundaries of lengths 
$\ell_n$, as illustrated in \figref{fig7} for the case $n=3$. Also in that 
case the counting can be done and the continuum limit taken.
The continuum definitions of the $n$-loop functions are
\bea
W(\ell_1,\ldots,\ell_n,V) \!\!&=& \int_{\ell_1,\ldots,\ell_n} 
\cD [g_{\a\b}] \;\del(A(g)-V), \label{2.10} \\
 W(\ell_1,\ldots,\ell_n,\Lam)\!\! &=& \int_{\ell_1,\ldots,\ell_n} 
\cD [g_{\a\b}] \;\e^{-\Lam A(g)}, \label{2.11}\\
W(\Lam^B_1,\ldots,\Lam^B_n,\Lam) \!\!&= &
\int\cD [g_{\a\b}] \;\e^{-\Lam A(g)-\sum_i \Lam_i^B \ell_i(g)}. \label{2.12}
\eea
Formally \rf{2.10} counts each continuous geometry 
(defined by an equivalence class of metrics $[g_{\a\b}(\xi)]$) with 
weight one. Eq.\ \rf{2.11} defines the partition function for universes
with fixed boundary lengths $\ell_i$ and with a cosmological constant $\Lam$.
Eq.\ \rf{2.12} defines the partition function for universes
with boundary cosmological constants $\Lam^B_i$ and bulk cosmological 
constant $\Lam$, i.e.\ the partition function where both the lengths of 
the boundaries and the size of the universe are allowed to fluctuate,
controlled by the various cosmological constants. From a 
``counting perspective'' one can view $ W(\Lam^B_1,\ldots,\Lam^B_n,\Lam)$
as the generating function for $W(\ell_1,\ldots,\ell_n,V)$, the 
number of continuous geometries with $n$ boundaries of lengths $\ell_i$.

Of course, to perform any real counting one has to introduce a regularization
such that one starts out with a finite number of geometries, and for this 
purpose the DT-formalism is perfect. As an example we can write
the regularized DT version of $W(\Lam^B,\Lam)$, i.e. the 1-loop function, as
\beq\label{2.13}
 W(z_1,g) =\frac{1}{z_1} 
\sum_{k,l_1} W(l_1,k) \; g^k z_1^{-l_1}, ~~~g=\e^{-\m},~z_1=\e^{\lam_1},
\eeq
such that $W(z_1,g)$ is the generation function for $W(l_1,k)$, 
the number of triangulations with $k$ triangles and a boundary
with $l_1$ links. As with most counting problems, it is easier first 
to find the generating function $W(z_1,g)$ and 
then by inverse (discrete) Laplace transformations to 
find the numbers $W(l_1,k)$. 

The result of this counting (\cite{ajm}, 
see \cite{book}, Chapter 4, for a review) 
is that after 
the continuum limit is taken, using the techniques of 
renormalization of the bare lattice cosmological constant $\mu$ and 
boundary cosmological constants $\lam_i$ (appearing in \rf{2.13}),
one obtains the expression
\beq\label{2.14}
W(\ell_1,\ldots,\ell_n,V)= V^{n-7/2} \sqrt{\ell_1 \cdots \ell_n}
\;\e^{-(\ell_1+\cdots+\ell_n)^2/V}.
\eeq
Starting out from the continuum Liouville theory the same 
result has been reproduced. In this sense the agreement 
shows that the DT lattice regularization works perfectly (and even 
allows one to perform certain analytic calculation with less effort
than using the continuum formulation, something very 
rare for a lattice regularization). It also gives 
additional confidence in the continuum Liouville calculations, which rely on 
certain bootstrap assumptions about conformal invariance.

\section{The fractal structure of 2d QG}

While eq.\ \rf{2.14} is an amazing formula, basically 
counting the number of continuous 2d geometries with 
the topology of a sphere with $n$ boundaries, it 
tells us little about the ``typical''  2d continuous geometry
one encounters in the path integral. In order to probe such a 
geometry we need some specific reference to distance. One 
could be worried that it makes no sense to talk about distance
in a theory of quantum gravity, i.e.\ a theory of fluctuating 
geometry, since it is precisely the geometry that
defines distance. However, the key message of the following
is that it {\it does} make sense to talk about geodesic distance 
even in a such a theory.

Let us define the two-point function $G(R;V)$ of 
geodesic distance $R$ for surfaces of fixed volume $V$ by
\bea\label{3.1}
\lefteqn{G(R;V)=} \\  
&&\int\! \cD [g]\!\int\! \cD_g \psi\; \e^{-S[g,\psi]}
\;\del\Big(A(g)\mi V\Big)  \!\!
\int\!\!dx\sqrt{g(x)}\!\!\int\!\! dy\sqrt{g(y)}  
\;\del(R \mi D_g(x,y)),  \nonumber
\eea
where $A(g) \equiv \int d^2x \sqrt{g(x)}$ and $D_g(x,y)$
denotes the geodesic distance between $x$ and $y$ in the geometry
defined by the metric $g_{\a\b}(x)$.
The defining formula \rf{3.1} is 
valid for any matter field $\psi$ coupled to 2d quantum gravity.
In principle it is also valid in a higher dimensional theory 
of quantum gravity provided one includes in $S[g,\psi]$ the 
Einstein action (or whatever one uses as the action). In two dimensions the 
Einstein action is topological and we may drop it.

It might be convenient not to keep $V$ fixed, but rather to
consider the two-point function for the ensemble of universes 
with a fixed cosmological constant $\Lam$, i.e.
\beq\label{3.2}
G(R;\Lam) = \int_0^\infty dV \; \e^{-\Lam \, V} \; G(R;V).
\eeq
These two-point functions probe the geometries in the following way.
Denote the ``area'' of a spherical shell at geodesic distance $R$
from point $x$ by
\beq\label{3.3}
S_V(x;R) = \int dy \sqrt{g(y)} \;\del(D_g(x,y)-R),
\eeq
which, of course, depends both on the chosen geometry $g_{\a\b}$ and 
the point $x$.
Let us denote the diffeomorphism invariant average of $S_V(x;R)$ by
\beq\label{3.3a}
S_V(R) = \frac{1}{V} \int dx \sqrt{g(x)} \; S_V(x;R).
\eeq
The quantum average of $S_V(R)$ over all geometries
is then related to $G(R;V)$ by
\beq\label{3.4}
\la S_V(R) \ra = \frac{1}{V Z(V)} \;G(R;V),
\eeq
where $Z_V$ is the corresponding partition function of 2d quantum 
gravity coupled to matter, i.e. the rhs of \rf{3.1} but 
with the integral (and integrand) over $x,y$ removed. For a smooth 
2d geometry we have 
\beq\label{3.5}
S_V(R) \sim R,~~~{\rm for}~~~R\ll V^{1/2},
\eeq
while in general we define the fractal dimension, or 
Hausdorff dimension, $d_h$ for the quantum average by 
\beq\label{3.6}
\la S_V(R) \ra \sim R^{d_h-1}~~~{\rm for} ~~~R\ll V^{1/d_h}.
\eeq
The partition function scales as $Z(V)\sim V^{\g(c) -3}$, where the
string susceptibility $\g(c)$ is a function of the central charge $c$ of 
the matter field coupled to the geometry and 
is known to be given by \cite{DDK,kazmig}
\beq\label{3.5b}
\g(c) = \frac{c-1-\sqrt{(c-1)(c-25)}}{12}.
\eeq
In the absence of matter 
fields, i.e.\ $c=0$, we have $\g=-1/2$, and the scaling is 
seen to agree with \rf{2.14}
for $n=0$. Therefore we can determine $d_h$ from the functional 
form of $G(R;V)$ or 
$G(R;\Lam)$. Remarkably, there is a simple and closed formula for 
$G(R;\Lam)$ for $c=0$, obtained again by counting 
triangulations, namely \cite{fractal}
\beq\label{3.7}  
G(R;\Lam) = \La^{3/4} \; 
\frac{\cosh( \sqrt[4]{\La} \; R)}{\sinh^3( \sqrt[4]{\La} \; R)}.
\eeq
This can be turned into an expression for $G(R;V)$ by an inverse Laplace 
transformation, which may plugged into \rf{3.6}, leading to
\beq\label{3.8}
\la S_V(R) \ra = R^3 F\left(\frac{R}{V^{1/4}}\right),~~~F(0) >0,
\eeq
where $F(x)$ is a hypergeometric function falling off for large
$x$ as $e^{-x^{4/3}}$. Note that, while $G(R;V)$ falls of faster 
than exponentially as a function of $R$, this is not possible
for $G(R;\Lam)$ because of arguments of subadditivity of the kind
already used for the two-point function of the bosonic string.  

Comparing \rf{3.8} to \rf{3.6}, we conclude that 2d 
continuous geometry is fractal with Hausdorff 
dimension $d_h=4$ \cite{kk,fractal1}.
This is in some sense similar to the situation for the free particle,
where one is summing over continuous path from $x$ to $y$ in $\mathbb{R}^d$. 
There a typical path is not a one-dimensional object, but is fractal with
$d_h=2$. The difference is that for the geometries we have no
embedding space $\mathbb{R}^d$ with respect to which we can define a distance. 
This makes it the more remarkable that one still has a concept of geodesic
distance that survives the averaging over all geometries. 

\begin{figure}
\centerline{\scalebox{0.4}{\rotatebox{0}{\includegraphics{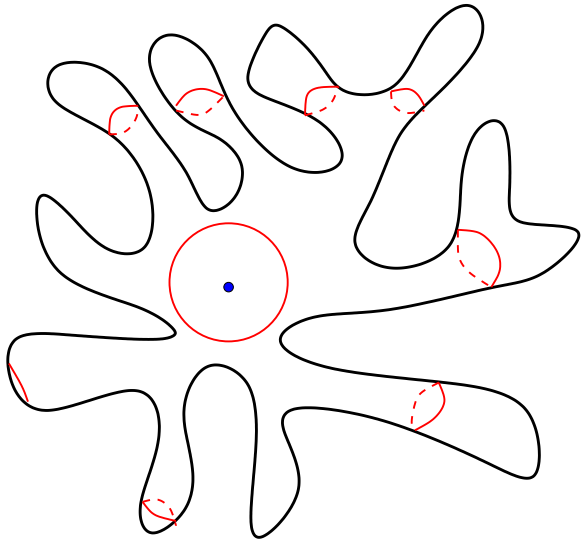}}}}
\caption{The fractal structure of a ``typical'' 2d geometry.}
\label{fig8}
\end{figure} 

How is it possible that $d_h=4$? The reason 
$d_h$ can be larger than 2 is that $S_V(x;R)$ is almost surely not 
connected, as is illustrated in \figref{fig8}.
In fact, one can show \cite{fractal} that the number of connected 
components of 
$S_V(x;R)$ with length $\ell$ between $\ell$ and $\ell+d\ell$ is given by 
\beq\label{3.9}
\rho_R(\ell) \propto \frac{1}{R^2} \left(y^{-5/2} + \oh y^{-3/2}+
\frac{14}{3} y^{-1/2}\right)
\, \e^{-y},~~~~y=\frac{\ell}{R^2},
\eeq  
in the limit $V\to\infty$.
Thus the number of components with small $\ell$ diverges for $\ell \to 0$.
Of course, in the DT formalism there is a cut-off in the sense that the
smallest loop length consists of a single link (of length $a$, the UV cut-off).
In the presence of such a cut-off \rf{3.9} leads to
\beq\label{3.10}
\la S_{V\to\infty}(R)\ra = \int_a^{\infty} d\ell \; \ell \; 
\rho_R(\ell) \propto
\frac{R^3}{\sqrt{a}},
\eeq 
again leading to the conclusion that $d_h=4$.

\subsection{The central charge different from zero}

For $c \neq 0$ (and $c\leq 1$) no detailed calculations exist
like the ones reported above. However, there exists a remarkable 
formula derived by Watabiki \cite{watabiki} for $d_h$ for any $c \leq 1$: 
\beq\label{3.11}
d_h(c) = 2 \frac{\sqrt{49-c}+\sqrt{25-c}}{\sqrt{25-c} +\sqrt{1-c}},
~~~~~~~~~
d_h(0) = 4,~~~~d_h( -\infty)=2.
\eeq
The formula was derived by applying scaling arguments, 
which we will briefly summarize, to diffusion on two-dimensional 
geometries in quantum Liouville theory.

Let $\Phi_n[g]$ be a functional of the metric which is 
invariant under diffeomorphisms
and assume that classically $\Phi_n[\lam g] = \lam^{-n} \Phi[g]$  for constant 
$\lam$. According to the KPZ relations the 
quantum average then satisfies \cite{kpz,DDK,watabiki}
\beq\label{3.12}
\la \Phi[g] \ra_{\lam V} = \lam^{-\a_{-n}/\a_1} \la \Phi[g] \ra_V,~~~~~
\a_n = \frac{2n}{1+\sqrt{\frac{25-c-24n}{25-c}}}
\eeq
One now applies this to the operator
\beq\label{3.13}
\Phi_{1}[g] = \int dx\sqrt{g} \;\left[\Del_g(x)\; 
\del_g(x,x_0)\right]_{x=x_0},~~~~
\Phi_1[\lam g]= \lam^{-1} \Phi_1[g], 
\eeq
which appears when we study diffusion on a smooth manifold 
with metric $g_{\m\n}$.
The diffusion kernel is 
\beq\label{3.14}
K(x,x_0;t)= \e^{t \Del_g} \,K(x,x_0;t),~~~~K(x,x_0;0) = \del_g(x,x_0).
\eeq
It has short distance behavior 
\beq\label{3.15}
K(x,x_0;t) \sim \frac{\e^{-D^2(x,x_0)/2t}}{t^{d/2}}\,(1+\cO(t)),~~~~~
 \la D(x,x_0;t)^2\ra \sim t +\cO(t^2).
\eeq
The {\it return probability} is defined in terms of the diffusion kernel as
\bea
P(t) &=& \frac{1}{V} \int dx \sqrt{g} \;K(x,x;t) \nonumber \\
&=& \frac{1}{V}
\int dx \sqrt{g} \left[(1+t \Del_g +\cdots) \;
\del_g(x-x_0)\right]_{x=x_0}\nonumber \\
&=& c +t\, \Phi_1[g] +O(t^2) \label{3.15b}
\eea
These equations are trivially correct for a smooth geometry $g_{\a\b}(x)$,
and they link the dimension of $\Phi_1[g]$ to the dimension of 
$ D(x,x_0)$:
\beq\label{3.16}
{\rm Dim} [ D(x,x_0)] = -\oh {\rm Dim}[\Phi[g]].
\eeq
Of course, this link is trivial in the sense that ${\rm Dim} [ D(x,x_0)]=1$
and ${\rm Dim}[\Phi_1[g]]=-2$ by construction. 
Watabiki now conjectured that \rf{3.16} survives the 
quantum averaging, where we know from \rf{3.12} how the dimension
of $\Phi_1[g]$ changes. Thus one obtains
\beq\label{3.17}
{\rm Dim} [\la D(x,x_0)\ra] = -\oh {\rm Dim}[\la \Phi[g]\ra] = 
-\frac{\a_{-1}}{\a_1},
\eeq
leading to \rf{3.11} if we declare that Dim[$V$] = 2, such that
\beq\label{3.18}
\la V \ra_R = R^{d_h},~~~~~~{\rm Dim} [R] = \frac{2}{d_h}. 
\eeq

\subsection{Is the Watabiki formula correct?}

One may be worried about the previous derivation of $d_h(c)$,
since the result implies that a typical spacetime is fractal,
while the basic relation used, namely \rf{3.15b}, is valid
only on smooth spacetimes. 
But not only that: numerical simulations \cite{qganddiff}
seem to show that the diffusion distance $R(t)$ scales
like $\la R^2(t)\ra \sim t^{2/d_h}$, rather than like in \rf{3.15}.
Anomalous diffusion is normal on fractal spacetimes, but it 
makes the Watabiki derivation problematic. Nevertheless, the predicted
$d_h(0)$ is clearly correct and it might be that $d_h(c)$ is also 
correct for $c\neq 0$. This is what we have tried to test 
using numerical methods to measure $d_h(c)$.  

\begin{figure}
\centerline{\scalebox{0.75}{\rotatebox{0}{\includegraphics{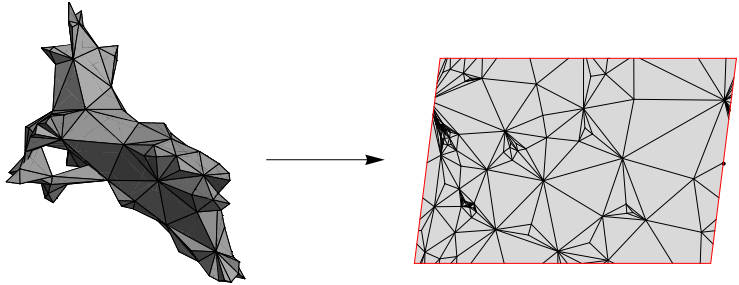}}}}
\caption{Example of a discrete analog of a harmonic map, used to map a triangulation of the torus consisting of equilateral 
triangles into the complex plane \cite{abb}.} 
\label{fig9}
\end{figure}

We have found it 
convenient to use 2d spacetimes with toroidal topology.
These have the virtue that their shortest non-contractible loop 
is automatically
a geodesic curve \cite{abbl}. Thus in the discretized case we only have 
to look for such loops. Further, the harmonic forms which 
are important tools for analytic manifolds 
have  very nice discretized analogies, and we 
can use the  these to construct a conformal mapping 
from the abstract triangulation to the complex plane \cite{abb,kawaigenus}.
We have shown an example of such a map in \figref{fig9}.

\begin{figure}
\centerline{\scalebox{0.58}{\rotatebox{0}{\includegraphics{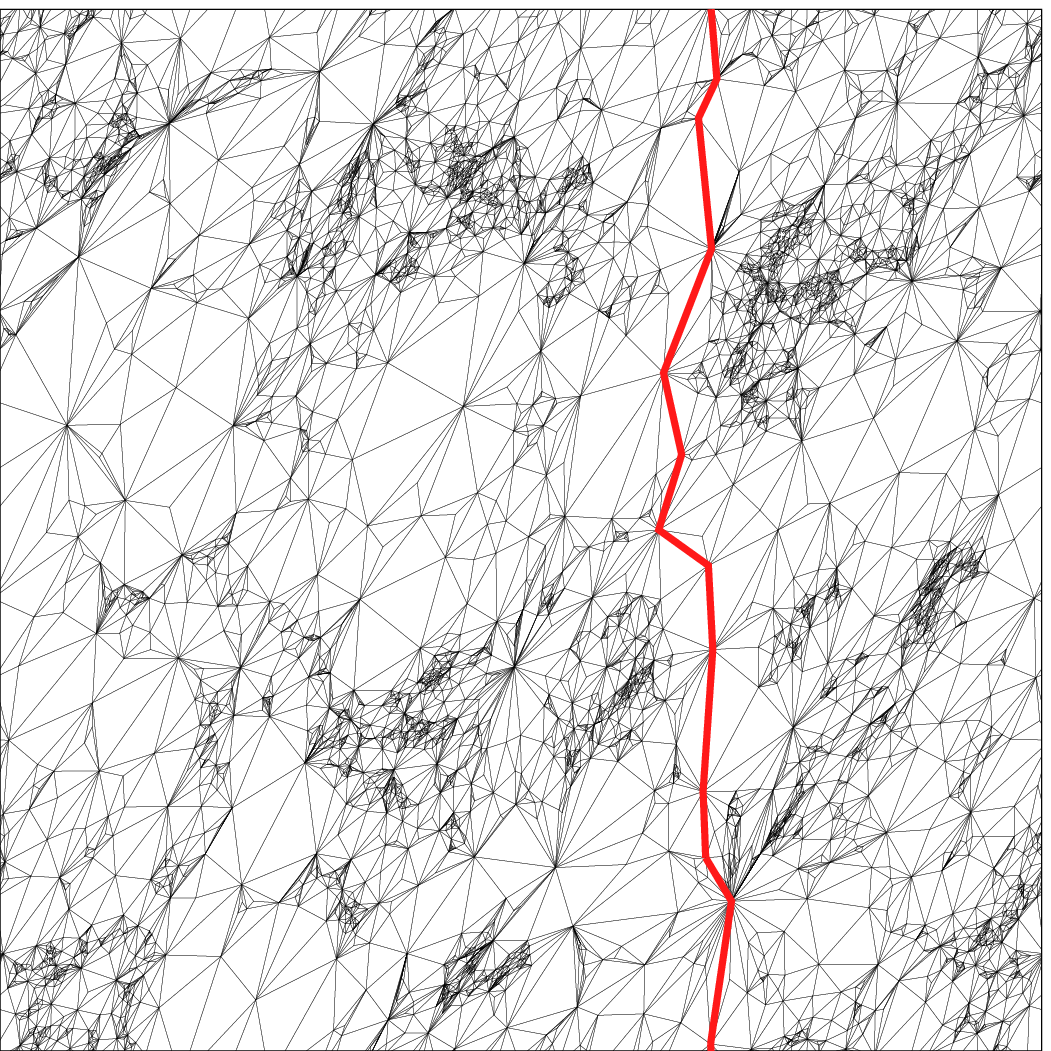}}}
~~\scalebox{0.58}{\rotatebox{0}{\includegraphics{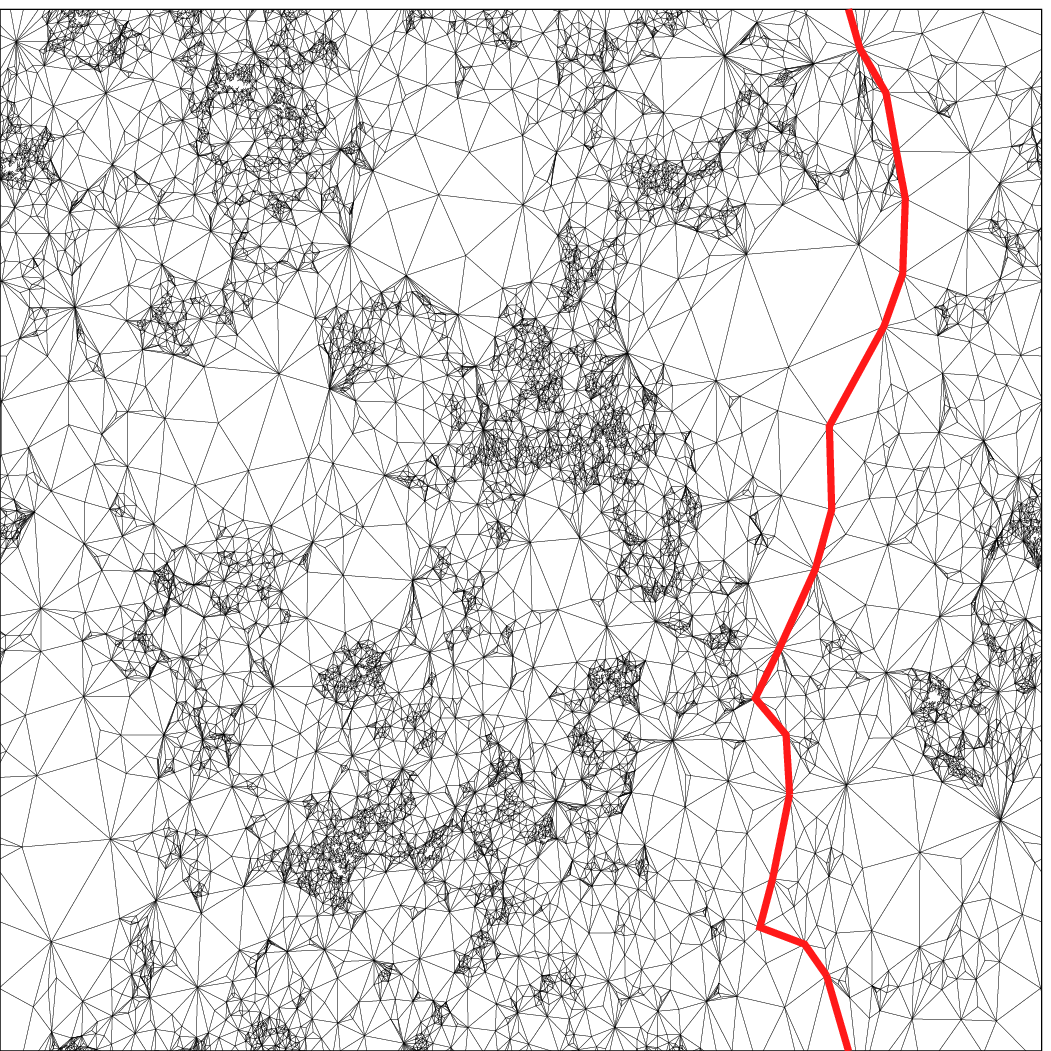}}}}
\caption{The left figure corresponds to $c= 0$, i.e. $d_h =4$, 
and the right figure to $c=-2$,  i.e.\ $d_h=3.56$. 
The shortest path non-contractible loop is shown in both cases \cite{abb}.}
\label{fig10}
\end{figure}

Since the shortest non-contractible loop is a geodesic we expect 
\beq\label{3.19}
\la L \ra_N \sim N^{1/d_h(c)} 
\eeq
An amazing qualitative test of this is shown in \figref{fig10},
where we use the harmonic map mentioned to map two abstract triangulations
corresponding to $c=0$ and $c=-2$ and 150000 triangles into the 
complex plane. Already just by looking at the figures one can basically 
verify qualitatively \rf{3.19}.

\begin{figure}
\centerline{\scalebox{1.05}{\rotatebox{0}{\includegraphics{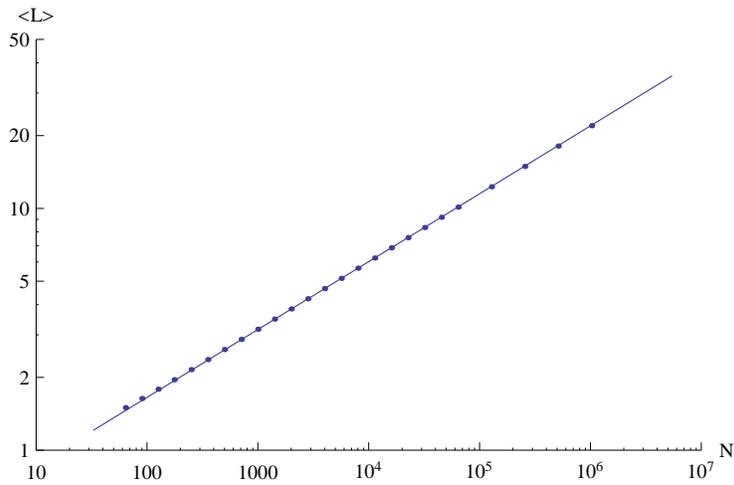}}}}
\caption{The numerical expectation value $\la L \ra_N$ of the 
length of the shortest non-contractible loop for 
triangulations of $N$ triangles (the error bars are too small to display).
The fit corresponds to $\la L \ra_N = 0.45\; N^{1/3.56}$ \cite{abb}.}
\label{fig11}
\end{figure}

A quantitative check of $\la L\ra_N \sim N^{1/d_h}$ for $c=-2$ is shown 
in \figref{fig11}, where we have averaged over many configurations 
for a fixed size $N$ of the triangulation, and performed the 
measurements of the shortest non-contractible loops for different 
sizes $N$. Formula \rf{3.11} seems very well
satisfied numerically for $c=-2$.

Recall that the partition function for the (regularized) 
bosonic string embedded in $d$ dimensions is given by eq.\ \rf{1.13}:
it can be viewed as a conformal field 
theory with central charge $c=d$ coupled to 2d quantum gravity. 
As we have seen, the theory degenerates into BP for $c>1$. However,
from \rf{1.13} it is clear that we can formally perform an
analytic continuation to $c<1$. A special case is $c=-2$ because then 
the triangulations are weighted precisely by the determinant of the graph Laplacian, which
can be represented as 
a sum over spanning trees of the given triangulations.
This fact was 
used in the numerical simulations reported above and allowed us
to sample very large triangulations and to 
obtain great numerical accuracy \cite{abb}. 

\begin{figure}
\centerline{\scalebox{0.35}{\rotatebox{0}{\includegraphics{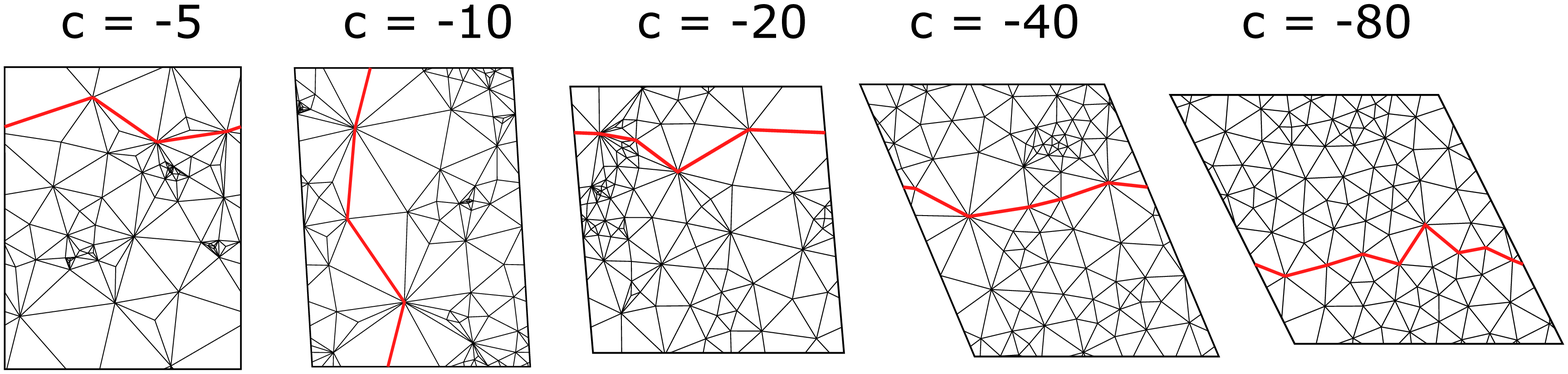}}}}
\caption{Qualitative agreement with  
\rf{3.11} for large negative $c$ \cite{absemiclas}.}
\label{fig12}
\end{figure}

More generally one can sample from the partition function for any fixed
real value $c$ by explicitly evaluating the determinant in a Monte Carlo
simulation \cite{absemiclas}.
This can, of course, only be done efficiently for relatively 
small triangulations.
However, it turns out that to study DT for large 
negative $c \ll -2$ and to obtain 
a qualitative verification of formula \rf{3.11}, 
one only requires such small triangulations.  
In particular, the formula tells us that 
$d_h \to 2$ for large negative $c$, indicating that nice 
smooth geometries should dominate in that limit. 
This is illustrated in \figref{fig12}. 

\begin{figure}
\centerline{\scalebox{0.35}{\rotatebox{0}{\includegraphics{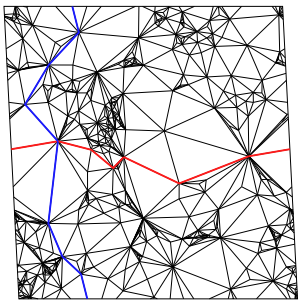}}}}
\caption{Example of two shortest, independent loops \cite{abhaus}.}
\label{fig13}
\end{figure}

The situation for $c > 0$ is  more difficult and until
recently numerical simulations could not really 
determine $d_h(c)$ properly for $c >0$. Matter correlation
functions gave agreement with Watabiki's formula, but 
geometric measurements agreed better with $d_h =4$ for $0<c<1$.
Recently simulations have been performed of DT on the torus coupled to 
the Ising model ($c=1/2$) and the 3-states Potts model ($c=4/5$) \cite{abhaus}.
In addition to the shortest non-contractible loop 
length $\ell_0$, also the length $\ell_1$ of the second
shortest independent loop was analyzed 
(see \figref{fig13}), yielding data with little discretization
``noise''.
The probability distributions for the lengths $\ell_i$ are expected,
 for large $N$, to be of the form
\beq\label{3.22}
P_N^{(i)}(\ell_i) = N^{1/d_h} F_i(x_i)~~~~x_i=\frac{\ell_i}{N^{1/d_h}}
\eeq
By measuring the distributions for various $N$'s and attempting to 
``collapse'' the distributions to the common, universal functions 
$F_i(x_i)$ we can determine $d_h$. Typically, one chooses
reference distributions, here chosen to be interpolations of 
the loop length distributions 
for $N=8000$, to which the data for the other system sizes is fitted. 
In \figref{fig13} the reference distributions 
$P_N(\ell_0)$ and $P_N(\ell_1)$ are plotted for both the 
Ising model and the 3-states Potts model.
It is seen that the second shortest loop distributions
contain less very short loops, which is probably why their lengths have 
less discretization effects and show better scaling. 
\begin{figure}[t]
\center{\scalebox{1.0}{\rotatebox{0}{\includegraphics{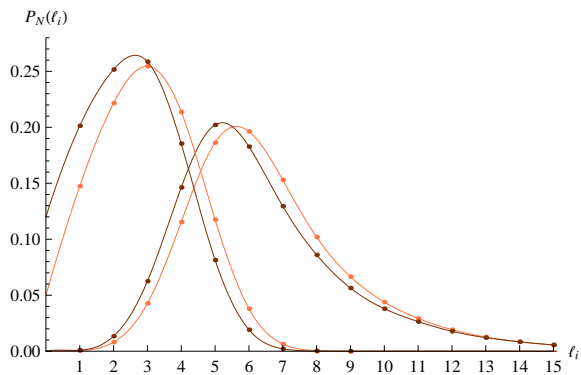}}}}
\caption{The reference distributions $P_N(\ell_0)$ (left) 
and $P_N(\ell_1)$ (right) for the Ising model (light curves) 
and the 3-states Potts models (dark curves) 
extracted from the data at $N=8000$ \cite{abhaus}.}
\label{fig14}
\end{figure}
The best fits of $d_h$ for the data are shown in \figref{fig15} and 
summarized in the following table.
\begin{figure}[t]
\center{\scalebox{0.8}{\rotatebox{0}{\includegraphics{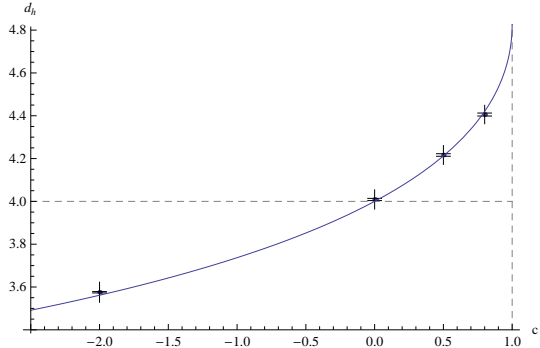}}}}
\caption{The results of high precision measurements of $d_h(c)$ \cite{abhaus}.
The shown curve is $d_h(c)$ as defined by eq.\ \rf{3.11}, and the 
measurements are for $c=-2,0,1/2,4/5$.}
\label{fig15}
\end{figure}
\begin{center}
\begin{tabular}{r|c|c}
$c$ & $d_h$ (by fit) & $d_h$ (theoretical) \\
\hline 
$-2$ & $3.575 \pm 0.003$ & $3.562$ \\
$0$ & $4.009 \pm 0.005$ & $4.000$ \\
$1/2$ & $4.217 \pm 0.006$ & $4.212$ \\
$4/5$ & $4.406 \pm 0.007$ & $4.421$ \\
\hline
\end{tabular}
\end{center}

\section{Matter correlation functions}

We have seen that the two-point functions $G(R;\Lam)$ and $G(R;V)$
are good probes of the quantum geometry of 2d spacetime
and allowed us to define the concept of an average geodesic 
distance. Also for matter correlators $\la \phi(x)\phi(y)\ra$ the first 
obvious question one can ask is whether it makes any sense 
to talk about such correlators as functions of distance,
and which distance should one use if we are integrating over 
all geometries? It is natural to define the diffeomorphism 
invariant matter correlators as the following generalization
of eq.\ \rf{3.1} for $G(R;V)$:
\bea\label{5.1}
\la \phi \phi (R) \ra_V &=&  
 \frac{1}{Z_V} \int\! \cD [g]\!\int\! \cD_g \phi\; \e^{-S[g,\phi]}
\;\del\Big(A(g)\mi V\Big)  \\ 
&&\int\!\!\int\!\! dx\,dy \;\frac{\sqrt{g(x)} \, \sqrt{g(y)}}{S_g(y,R)\; V}  
\; \phi(x) \,\phi(y) \;\del(R \mi D_g(x,y)).  \nonumber
\eea  
It is a non-local definition of a matter correlator, but  
there exists no diffeomorphism invariant local definition.

Assume we consider  a conformal field theory in flat spacetime
and let $\phi(x)$ be a primary operator with scaling dimension $\Del_0$.
We thus have the following behavior of the $\phi\mi \phi$ correlator
\beq\label{5.2}
\la \phi(x)\phi(y)\ra \sim |x-y|^{-2 \Del_0}.
\eeq
If we take the quantum average as in \rf{5.1} the geodesic 
distance $R$ scales anomalously and we expect for dimensional 
reasons that $|x-y|^{-2}$ is replaced by $R^{-d_h}$. However, 
we also know from KPZ scaling that the scaling dimension $\Del_0$ 
of $\phi$ will be changed after coupling to 2d quantum gravity
such that  
\beq\label{5.3}
\Del(c) = 2 \;\frac{\sqrt{1-c + 12 \Del_0}-\sqrt{1-c}}{\sqrt{25-c}-\sqrt{1-c}},
~~~~{\rm KPZ-DDK~scaling},
\eeq
where one observes that $c\to -\infty$ implies $\Del(c) \to \Del_0$ 
in agreement with the earlier observation that Watabiki's formula 
shows that  $d_h (c) \to 2$ for $c\to -\infty$.
For a finite spacetime volume $V$ we finally expect a behavior
\beq\label{5.4}
\la \phi \phi(R)\ra_V \sim \frac{1}{R^{d_h \Del}} \; \,
g_\phi\left(\frac{R}{V^{1/d_h}}\right),
\eeq
which alternatively can be written as 
\beq\label{5.5}
\la \phi \phi(R)\ra_V \sim V^{-\Del}\;\; \frac{g_{\phi}(x)}{x^{d_h \Del}},
~~~~~~x=\frac{R}{V^{1/d_h}}
\eeq
For a given conformal field theory  $g_\phi(x)$ 
is a universal finite size function with 
$g_\phi(0)= const. > 0$ 
and $g_\phi(x)$ falling  of at least exponentially fast for $x > 1$.

The formula \rf{5.5} is convenient to use in the DT regularization
where $V \sim N_T$ and the geodesic distance $R \sim \ell$, 
the link distance between two vertices:
\beq\label{5.6}
\la \phi \phi(\ell)\ra_N \sim N^{-\Del}\;\; \frac{g_\phi(x)}{x^{d_h \Del}}
~~~~~~x=\frac{\ell}{N^{1/d_h}}-
\eeq
We note that eq.\ \rf{5.6} has for form of a standard finite size 
scaling relation. One can thus apply the formula to the Ising model 
or 3-states Potts model and measure the spin-spin correlation
functions for various values of $N$. Collapsing these correlation
functions to  universal functions $g_\phi(x)$ for either the 
Ising model ($c=1/2)$ or the 3-states Potts model $(c=4/5)$ 
allow us to determine 
both $d_h(c)$ and $\Del(c)$ for these values of $c$. 
One finds a $d_h(c)$ in agreement 
with watabiki's formula as mentioned earlier (but not with 
the same precision as with the method described in the last section),
and one finds a $\Del(c)$ in good agreement with the KPZ formula \rf{5.3}
\cite{aa}. The result of collapsing the data to a (best possible) 
universal function $g_\phi(x)$ is shown in \figref{fig10} for the Ising
model. It works very well for an impressive range of lattice sizes. Remarkably,
finite size scaling works even better on the DT-ensemble of lattices
than on a fixed lattice. Somehow the random lattices average out 
finite lattice artifacts.  
\begin{figure}[h]
\centerline{\scalebox{0.35}{\rotatebox{0}{\includegraphics{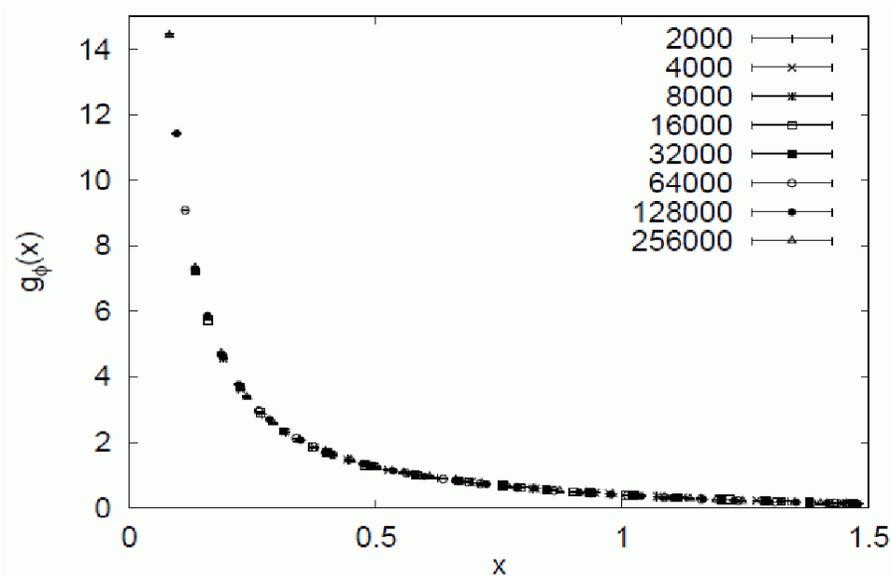}}}}
\caption{The Ising spin correlation functions collapsed to the universal 
function $g_\phi(x)$ for  the range of lattice
sizes listed (measurements done for the Ising model at its 
annealed average critical point).} 
\end{figure}

\section{Conclusions}

Two-dimensional quantum gravity is a nice playground
for testing to what extent it makes sense 
to talk about non-trivial diffeomorphism invariant 
theories of fluctuating geometry. We have here focused on the very simplest 
question: if one integrates over the fluctuating geometries
as one should do in a path integral representation 
of a quantum theory, how can one at all talk about 
concepts like distances and correlation functions falling 
off with this distance. In this context 2d quantum gravity
is the perfect theory for such tests. It has no propagating 
gravitational degrees of freedom, but 
it is maximally {\it quantum}, the reason precisely being 
that the Einstein action in two dimensions is trivial. 
Every geometry carries therefore the same weight in the path integral,
as exemplified by eq.\ \rf{2.10}, i.e.\ formally it 
corresponds to a $\hbar \to \infty$ limit.
If we want to study ordinary field theories
(like conformal field theories) and not just esoteric 
topological field theories, we cannot 
avoid the clash between the integration over all 
geometries and the need to have some concept of  
distance. However, as we have seen some aspects of geodesic distance 
remarkably survive
this quantum average over geometries, despite the fact
that geodesic distance is a awfully non-local notion.
Although the geodesic distance picks up an anomalous dimension
due to quantum fluctuations, it maintains its role as the distance 
which can be used in the correlators between fields. 

In higher dimensions there might not exist a well-defined,
stand-alone theory of quantum gravity. The UV problems 
for such a theory might be too severe. This question is 
still up in the air, and it might well be that the metric 
degrees of freedom we have in classical GR are not the 
fundamental degrees of freedom one should use in the UV regime.
However, the studies reported here show that conceptually 
there seems to be no problem with a theory of ``fluctuating'' 
geometries per se and even in the most radical such one, namely 
2d quantum gravity, one can maintain many of the concepts we 
know from flat spacetime.

\section*{Acknowledgments}\label{app}

The authors acknowledge support from the ERC-Advance grant 291092,
``Exploring the Quantum Universe'' (EQU). JA acknowledges support 
of FNU, the Free Danish Research Council, from the grant 
``quantum gravity and the role of black holes''.
JA thanks his collaborators K. Anagnostopoulos, B. Durhuus, T.Jonsson,
J. Jurkiewicz and Y. Watabiki for many discussions on the topics 
covered here. They cannot be blamed for any mistakes (conceptional 
or otherwise) in this article.


\begin{thebibliography}{10}
\providecommand*{\bibinfo}[2]{#2}
\providecommand*{\eprint}[1]{#1}
\providecommand*{\url}[1]{#1}


\bibitem{david}
  F.~David,
  Nucl.\ Phys.\  {\bf B257 } (1985)  45.\\
A.~Billoire and F.~David,
Phys.\ Lett.\  B\ {\bf 168} (1986) 279-283.

\bibitem{kkm}
  V.~A.~Kazakov, A.~A.~Migdal, I.~K.~Kostov,
  Phys.\ Lett.\  {\bf B157 } (1985)  295-300.\\
D.V.~Boulatov, V.A.~Kazakov, I.K.~Kostov and A.A.~Migdal,
Nucl.\ Phys.\  B\ {\bf 275} (1986) 641-686.


\bibitem{adf}
J.~Ambjorn, B.~Durhuus and J.~Fr\"ohlich,
Nucl.\ Phys.\  B\ {\bf 257} (1985) 433-449;\\
J.~Ambjorn, B.~Durhuus, J.~Fr\"ohlich and P.~Orland,
Nucl.\ Phys.\  B\ {\bf 270} (1986) 457-482.

  

\bibitem{book}
  J.~Ambjorn, B.~Durhuus, T.~Jonsson,
  Cambridge, UK: Univ. Pr., 1997. (Cambridge Monographs in 
Mathematical Physics). 363 p.


\bibitem{ad}
  J.~Ambjorn and B.~Durhuus,
  Phys.\ Lett.\ B {\bf 188} (1987) 253.



\bibitem{kazakov}
  V.~A.~Kazakov,
  Phys.\ Lett.\  {A\ 119} (1986)  140-144.


\bibitem{ajm}
  J.~Ambjorn, J.~Jurkiewicz and Y.~.M.~Makeenko,
  Phys.\ Lett.\ B {\bf 251} (1990) 517.


\bibitem{DDK}
  F.~David,
  Mod.\ Phys.\ Lett.\  {\bf A3 } (1988)  1651.\\
  J.~Distler, H.~Kawai,
  Nucl.\ Phys.\  {\bf B321 } (1989)  509.
  

\bibitem{kazmig}
  V.~A.~Kazakov and A.~A.~Migdal, Nucl.\ Phys.\ B {\bf 311} (1989) 171.


\bibitem{kk}
  N.~Kawamoto, V.~A.~Kazakov, Y.~Saeki, Y.~Watabiki,
  Phys.\ Rev.\ Lett.\  {\bf 68 } (1992)  2113-2116.
  

\bibitem{fractal}
  H.~Kawai, N.~Kawamoto, T.~Mogami, Y.~Watabiki,
  Phys.\ Lett.\  {\bf B306 } (1993)  19-26.
  [hep-th/9302133].\\
  J.~Ambjorn, Y.~Watabiki,
  Nucl.\ Phys.\  {\bf B445 } (1995)  129-144.
  [hep-th/9501049].

\bibitem{fractal1}
  J.~Ambjorn, J.~Jurkiewicz, Y.~Watabiki,
  Nucl.\ Phys.\  {\bf B454 } (1995)  313-342.
  [hep-lat/9507014].


\bibitem{watabiki}
  Y.~Watabiki,
  Prog.\ Theor.\ Phys.\ Suppl.\  {\bf 114 } (1993)  1-17.


\bibitem{kpz}
  V.~G.~Knizhnik, A.~M.~Polyakov, A.~A.~Zamolodchikov, 
Mod.\ Phys.\ Lett.\ A{\bf 3} (1988) 819


\bibitem{qganddiff}
  J.~Ambjorn, K.~N.~Anagnostopoulos, T.~Ichihara, L.~Jensen, Y.~Watabiki,
  JHEP {\bf 11 } (1998)  022
  [hep-lat/9808027].

\bibitem{abb}
  J.~Ambjorn, J.~Barkley and T.~G.~Budd,
  Nucl.\ Phys.\ B {\bf 858} (2012) 267
  [arXiv:1110.4649 [hep-th]].


\bibitem{kawaigenus}
  H.~Kawai, N.~Tsuda, T.~Yukawa,
  Phys.\ Lett.\  {\bf B351 } (1995)  162-168.
  [hep-th/9503052].
  Nucl.\ Phys.\ Proc.\ Suppl.\  {\bf 47 } (1996)  653-656.
  [hep-lat/9512014].
  H.~Kawai, N.~Tsuda, T.~Yukawa,
  Nucl.\ Phys.\ Proc.\ Suppl.\  {\bf 53 } (1997)  777-779.
  [hep-lat/9609002].


\bibitem{abbl}
  J.~Ambjorn, J.~Barkley, T.~Budd and R.~Loll,
  Phys.\ Lett.\ B {\bf 706} (2011) 86
  [arXiv:1110.3998 [hep-th]].
  
\bibitem{absemiclas}
  J.~Ambjorn and T.~G.~Budd,
  Phys.\ Lett.\ B {\bf 718} (2012) 200
  [arXiv:1110.5158 [hep-th]].
   
\bibitem{abhaus}
  J.~Ambjorn and T.~G.~Budd,
  Phys.\ Lett.\ B {\bf 724} (2013) 328
  [arXiv:1305.3674 [hep-th]].

\bibitem{aa}
  J.~Ambjorn and K.~N.~Anagnostopoulos,
  Nucl.\ Phys.\ B {\bf 497} (1997) 445
  [hep-lat/9701006].


\end{thebibliography}
\end{document}